\documentclass[twocolumn,aps,pre,floatfix]{revtex4}
\usepackage{graphicx,color,amsmath}

\newcommand{\bzpot}{%
\begin{figure}[htbp]
\begin{center}
\includegraphics[width=3.2in]{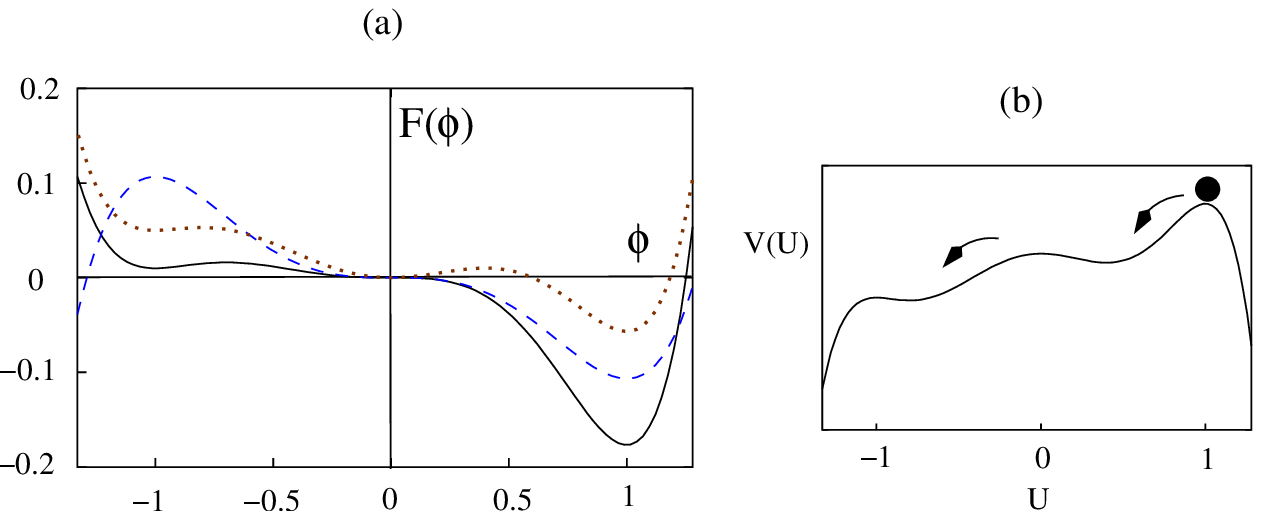}
\end{center}
\caption{(a)The Landau function $F(\phi)$ as a function of $\phi$.  
  In all cases, $\phi=1$ is the stable state, Z-DNA, $\phi=-1$ 
  represents an unstable (dashed line) or metastable (solid and dotted 
  lines) state, B-DNA while $\phi=0$ is a quadratically unstable (solid 
  and dashed lines) or metastable (dotted line) state, denatured state.  
  The three cases I, II and III in the text correspond to dotted, solid 
  and dashed lines.  (b)Potential $V(U)=-F(U)$ for the particle-on-a-hill 
  analogy.
}
\label{fig:1}
\end{figure}
}

\newcommand{\betavelotwo}{%
\begin{figure}[htbp]
\begin{center}
\includegraphics[width=3.3in]{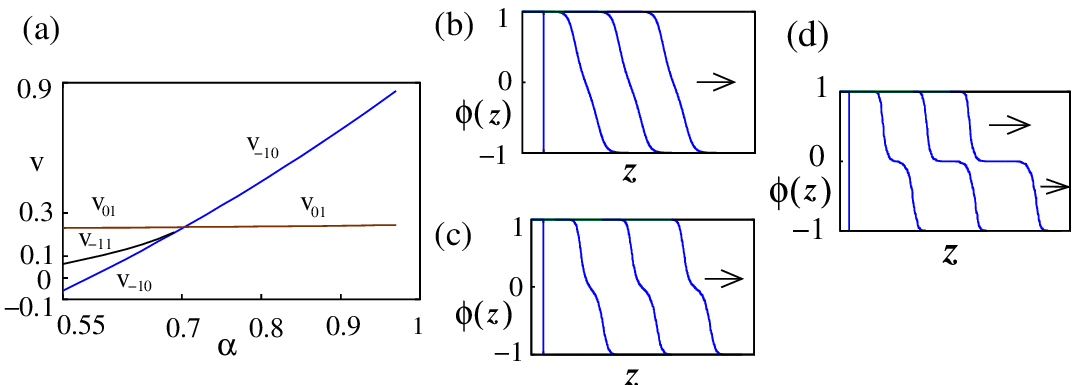}
\end{center}
\caption{(a)Plot of velocity {\it vs} $\alpha$ for a fixed value of
  $\beta=0.45$.  Three velocities meet at a common point at 
  $\alpha_c\left (\beta\right )$.
  The remaining three figures($\phi$ {\it vs} $z$) represent the time 
  evolution of the front(or fronts). 
  (b)A single front for $\alpha=0.6<\alpha_c\left (\beta\right )$.  
  (c)A single front for $\alpha=0.7$ near $\alpha_c\left (\beta\right )$ 
   with a signature of the width widening but no ``0'' phase. 
  (d)For $\alpha=0.72>\alpha_c\left (\beta\right )$ single front splits 
  into two fronts.  
}
\label{fig:2}
\end{figure}
}

\newcommand{\phasedia}{%
\begin{figure}[htbp]
\begin{center}
\includegraphics[width=2.0in]{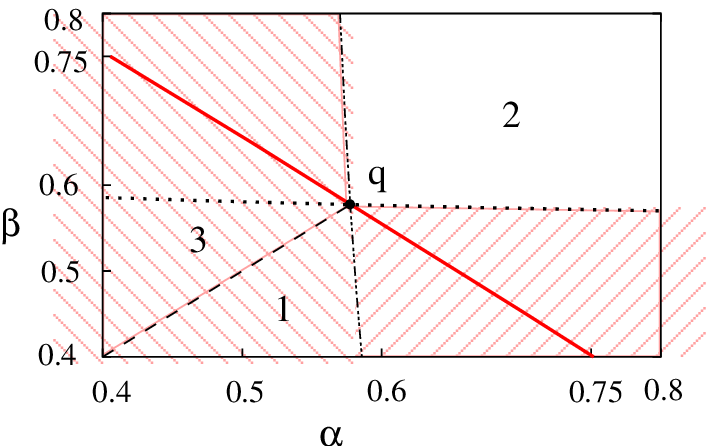}
\end{center}
\caption{Dynamic phases in a plot of $\alpha$ {\it vs} $\beta$, the
  boundary(solid line) being given by $\alpha=\alpha_c\left
    (\beta\right )$.  In the region below the boundary line, a single
  front between $-1$ to $1$(big front) propagates without splitting.
  In the region above the boundary line the front between $-1$ to $+1$
  splits into two (small) fronts.  Z, ``0'' and B are the
    stable states in regions 1, 2 and 3 respectively.  The dotted
  line corresponds to $v_{01}=0$, while the dash-dotted line
    to $v_{-10}=0$.  The split fronts move away from each other in
    region 2, both towards right in 1 and both towards left in 3, as
    per the chosen boundary conditions.  The big front has zero
    velocity on the $\alpha=\beta$ line and the diagram is symmetric
    around this line. Point q represents the equilibrium point, where
  three states have the same free energy.}
\label{fig:3}
\end{figure}
}

\newcommand{\width}{%
\begin{figure}[htbp]
\begin{center}
(a)\includegraphics[width=1.25in]{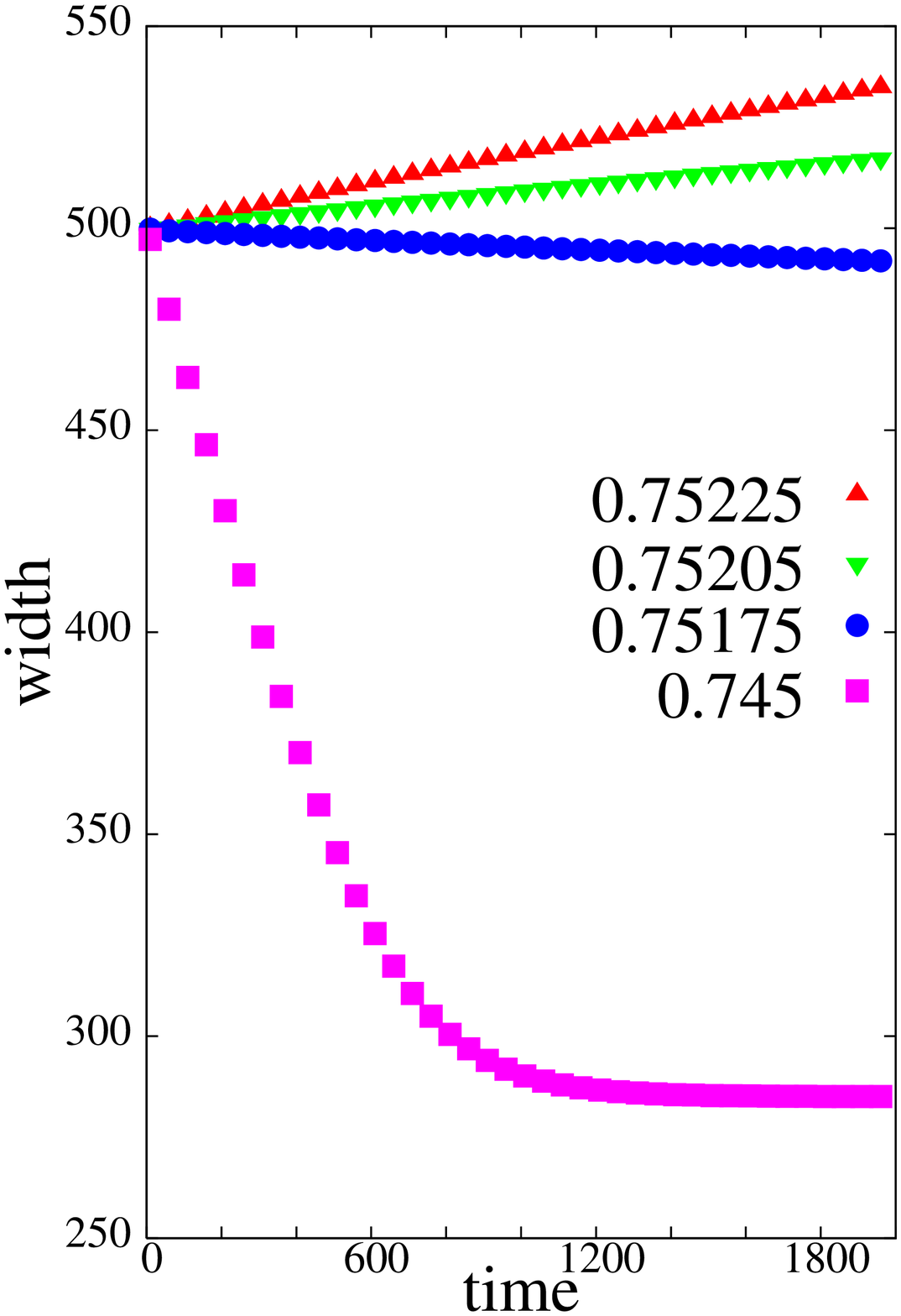}
(b)\includegraphics[width=1.65in]{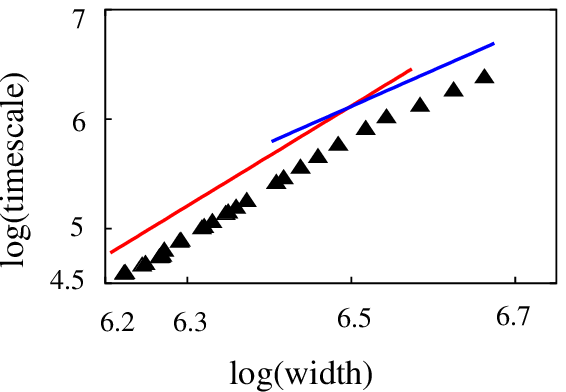}
\end{center}

\caption{(a) Time evolutions of the width (in arbitrary units) of
  the front are shown for different values of $\alpha$ keeping
  $\beta=0.4$ fixed.  The initial front had a stretch of ``0''
  phase which decays for $\alpha<\alpha_c\left (\beta\right )$ but
  grows linearly for $\alpha>\alpha_c\left (\beta\right )$.  The
  timescale to reach saturation increases as
  $\alpha\to \alpha_c\left (\beta\right )-$.  For this case
  $\alpha_c\left (\beta\right )\approx 0.752$.
  (b) Log-Log plot of width versus time scale for
  $\alpha<\alpha_c\left (\beta\right )$.  Two solid line slopes are shown 
  far from and near $\alpha_c\left (\beta\right )$.
}

\label{fig:4}
\end{figure}
}

\begin{document}
\title{Dynamic phase transition in the conversion of B-DNA to Z-DNA}
\author{Jaya Maji}
\email{jayamaji@iopb.res.in} 
\author{Somendra M. Bhattacharjee}
\email{somen@iopb.res.in} 
\affiliation{Institute of Physics, Bhubaneswar-751005, India   
} 
\begin{abstract}
  The long time dynamics of the conformational transition from
  B-DNA to Z-DNA is shown to undergo a dynamic phase transition.
  We obtained the dynamic phase diagram for the stability of the
  front separating B and Z.  The instability in this front results
  in two split fronts moving with different velocities.  Hence,
  depending on the system parameters a denatured state may develop
  dynamically eventhough it is thermodynamically forbidden.  This
  resolves the current controversies on the transition mechanism
  of the B-DNA to Z-DNA.
\end{abstract}
\date{\today}
\maketitle

\section{INTRODUCTION}
The most common form of DNA found under normal physiological low salt
conditions is the B-DNA, the well known double helix with a right handed 
helicity.   Quite surprisingly, the first DNA structure to be solved by 
X-ray crystallography turned out to be a left handed zig-zag form called 
the Z-DNA\cite{wang,bzrev,sinden}.  
This Z-DNA can be stabilized in vitro in presence of high salt 
concentration, cations or negative super-coiling.  Although the Z form 
is transient in vivo due to the lack of friendly environment, still the 
B-Z conversion is relevant in poxviruses\cite{rich}, and Alzeimer's 
disease\cite{alzeimer}.
Apart from the inversion of helicity, the Z-DNA has a repeat unit of 2
base pairs compared to one for B.  Thus a conformational
transformation from the B to the Z form takes place as, e.g., the
ionic concentration or super-coiling is changed.  The B-Z transition is
first order in nature\cite{wang,harvey,lim,kasten,ivanov}.

As the base pairs and a subset of backbone sugar rings have to flip to 
execute the B-Z transition involving changes in helical chirality, the 
dynamics offers intriguing possibilities\cite{bzrev}.
Only recently methods have been developed to explore the dynamics 
in single DNA as opposed to earlier studies in solutions,  
though with conflicting results.  
In Ref.~\cite{b} the B-Z conformational transformation for a short 15 base 
pair GT (non-Watson-Crick wobble base pair) DNA wrapped on a single walled 
carbon nanotube was monitored as a function of time by the addition 
of counter-ions.  
The nanotube helped in identifying the phases via accurate measurements 
of the band-gap in a simpler geometry.  This transition is completely
reversible and is thermodynamically identical to the transition seen
in the absence of the nanotube.  {\it The results seem to indicate the
formation of a denatured} DNA {\it during the transformation}, eventhough a
denatured state under the experimental conditions is not possible
thermodynamically.  A different single molecule experiment studied the
transition under a tension and negative super-helicity by combining
FRET with magnetic tweezers\cite{lee}.  This experiment on an
effectively (GC)$_{11}$ DNA (i.e. 22 bases) {\it seems to favour a
single interface between} B {\it and} Z {\it without any denatured bubble}.

One can characterize the B-Z transformation by a growth of a suitable
domain over the bulk of DNA.  In any such scenario, the B-Z interface, 
the separator between the two chiral phases, plays an important role.
The equilibrium interface has been characterized structurally and from
other studies.  The structure of a short oligomer in presence of a
Z-DNA binding protein at 2.6A resolution indicates broken base pairs 
separating the B and the Z phases.  The protein acting as an external 
source inducing the transition is expected to produce a sharp 
interface\cite{ha}.  A very ingenious way of studying the interface is 
to use mirror DNA\cite{urata,vichier}, though it cannot be used for 
chirality changing transition.  Interfacial studies and melting of short 
B-B* oligomers, where B* is the enantiomer of B, show that the junction 
mimics the B-Z junction, and the interface broadens as the melting point 
is reached.  In contrast to these equilibrium cases, the nature of the 
interface during the transition depends on the nature of the transition 
mechanism\cite{lim,kasten}.  Several such schemes are in vogue and 
discussed in detail in Ref.~\cite{bzrev}.
The two main competing hypothesis for the B to Z transition mechanism
are the following.  It is either via (1) the base pair separation
followed by base pair flipping\cite{wang}, or (2) the base pair
flipping without any base pair separation\cite{harvey}.  In the first
case there could be a denatured intermediate state while in the second
there could a Z type but following the standard Watson-Crick base
pairing (ZWC-DNA).

The apparently contradictory results from the two single molecule
experiments and the controversy associated with the B-Z transition mechanism 
in general, motivate us to study a coarse-grained thermodynamic model for
the dynamics. This implicitly requires infinitely long chains, since
small length DNAs or oligo-nucleotides do not show any proper
thermodynamic phase transition.  Inducing the B to Z transition is
tantamount to a lowering of the free energy of Z compared to B making
Z the most preferable state.  We in our formulation use the
simplification of the single molecule experiments to restrict the
geometry to one-dimension only.  The boundaries of the long chain are
maintained in the two states so that the new structure develops from
one side.  In such a problem the dynamics of the transition produces a
steady state with uniformly moving front (or fronts).  An investigation 
of the various types of fronts would help us answer the question of any 
dynamic generation of thermodynamically forbidden state.
\bzpot
\section{MODEL}
Our model consists of three states B, the denatured state and Z to be
represented by a parameter $\phi=-1, 0, 1$.  The space time coordinates
$z, t$ are taken to be continuous.  It is a one dimensional problem where
$\phi(z,t)$ describes the state of the coarse-grained base-pair at index
$z$ along the DNA.  For the B-Z transition, we take $\phi=-1$(B
state) to be unstable(or metastable) which is getting invaded by the
stable state at $\phi=1$(Z state).  
We study this phenomenon through a Landau free energy 
$F(\phi)$  taken as a sixth order polynomial
with the coefficients chosen to have extrema at $\phi=0,\pm 1$.  This is
ensured by choosing the thermodynamic force $f(\phi)$ as 
\begin{equation}\label{Eq:1}
f(\phi)=-\frac{dF(\phi)}{d\phi}=\phi(\phi+\alpha)(\phi-\beta)(1-\phi)(1+\phi),
\end{equation} 
where $\alpha, \beta>0$ are constants, whose values are system specific.  
Needless to say, the relative stability of the three phases can be adjusted 
by $\alpha, \beta$.  The Landau Ginzburg free energy is taken as 
\begin{equation}\label{Eq:2} {\cal H}(\phi)=\int
  dz\left[\frac{D}{2}\left(\frac{\partial \phi}{\partial
        z}\right)^2+{F}(\phi)\right],
\end{equation}
where $D>0$ is the elastic constant.  $D$-term allows inhomogeneity, e.g., 
at the interface between two phases.  The three homogeneous phases are given 
by the minima of the Landau free energy $F(\phi)$.  The dynamics is governed 
by the non linear diffusion equation
\begin{equation}\label{Eq:3}
\frac{\partial{\phi}}{\partial{t}}=D\frac{\partial^2{\phi}}{\partial{z^2}}+
f(\phi),
\end{equation}
derived from Eq.~{\ref{Eq:2}} in the overdamped limit.  The friction 
coefficient has been absorbed in the definition of time $t$.       
The geometry to be considered is such that the B state is on one side 
and the Z state on the other with the front moving towards the unstable 
state.  For the B-Z case, this is ensured by the boundary conditions 
$$\phi(z\to -\infty)=1,\quad \phi(z\to\infty)=-1$$ for Eq.~{\ref{Eq:3}} for 
all time.  A few other boundary conditions are considered too.
The three generic cases obtained by fixing $\alpha$ and $\beta$
are the following(see Fig.~{\ref{fig:1}}a)
\begin{itemize}
\item {Case I : While quenching to the stable state, Z, state B remains
      in a metastable state while the denatured state $\phi=0$ is also
      metastable.  Since the barriers are somewhere in between $\phi=-1$ 
      and $\phi=1$, we have $0\le\alpha, \beta<1$.}
\item {Case II : The metastable state(B-DNA) sees a barrier somewhere 
      inbetween $-1$ to $0$, while the denatured state is quadratically 
      unstable state.  This case is for $0<\alpha<1$, and $\beta=0$.}
\item {Case III : Unstable B state quenched into stable Z while the 
      denatured state remains in a quadratically unstable state (i.e., 
      without facing any barrier). This happens when $\alpha>1$ and 
      $\beta=0$.}
\end{itemize}
To be noted that cases I and II are similar to the free energy landscape 
obtained in Ref.~\cite{leeJPC} as the potential of the mean force obtained 
from molecular dynamics. 

The diffusive like term in Eq.~{\ref{Eq:3}} coming from the elastic part 
of Eq.~{\ref{Eq:2}} tends to smoothen out any  inhomogeneity while the 
driving force $f(\phi)$ tends to favour the stable state whenever there 
is any inhomogeneity.  The combined effect of the diffusion like spreading 
and the selection of one phase by the drive leads to a steady state where 
the interface shows a uniform motion and takes a shape which is not 
necessarily the equilibrium shape\cite{fkpp}.  
Based on the Fisher-Kolmogorov(F-K) idea, the traveling wave solution 
$\phi(z,t)=U(z-vt)$ can be used to rewrite Eq.~{\ref{Eq:3}} as  
\begin{equation}\label{Eq:4}
\frac{d^2U}{d\tau^2}+v \frac{dU}{d\tau} +f(U)=0,\quad(\tau=z-vt),
\end{equation}
where $v$ the velocity of the front is to be determined.  
The interface which we are studying is between $\phi=+1$ and $\phi=-1$
states.  Eq.~{\ref{Eq:4}} can be interpreted as the motion of a particle 
moving in a potential $V=-F(U)$(Fig.~{\ref{fig:1}}b) starting at the hill 
at $U=+1$ at time $\tau=-\infty$ just reaching the other hill at $U=-1$ 
at time $\tau=+\infty$ losing energy due to ``friction'' $v$.
\betavelotwo
For a given potential, such a motion is possible only for particular
values of $v$ and that velocity is the selected velocity of the front.
However, it is also possible that the particle spends an infinite
amount of time in the intermediate state so that the descent from
$U=+1$ to $U=0$ and the descent from $U=0$ to $U=-1$ are independent
requiring two different friction coefficients.  
The physical picture that emerges is that the stable state moves towards 
the unstable state, and the propagating front will have a time independent 
shape and a constant velocity $v$.  However in some situations, the 
initial big front separating the two phases $\phi =\pm 1$ splits into 
two, one front between $\phi=-1$ and $\phi=0$, while the other one between 
$\phi=0$ and $\phi=1$.  The two smaller fronts move with different shapes 
and speeds $v_{-10}, v_{01}$.  
The $\phi=0$ state may then get dynamically generated.  Consequently one 
may see the development of the denatured state.  The less preferable state 
will eventually be devoured by the stable state completing the transition 
from B to Z-DNA.

\section{Dynamic phase diagram-Numerical and perturbative approach}
The velocity of the front has been determined by numerical analysis for 
different boundary conditions like 
(a)~$\phi(-\infty,t)=1$, $\phi(\infty,t)=-1$ for the B-Z front, 
(b)~$\phi(-\infty,t)=1$, $\phi(\infty,t)=0$ for a front between 
Z and the denatured state, (c)~$\phi(-\infty,t)=0$, $\phi(\infty,t)=-1$ 
for a front between B and the denatured state.    
The initial($t=0$) interface of width $w$ is located at $z=z_0$ and a 
Crank Nicolson method is used to evolve the nonlinear diffusion equation.  
Once a steady state is reached, the velocity is determined by locating the
positions at which  $\phi=\pm .5$, and $\phi=0$ as appropriate.
In the case of the split front, only the velocity $v_{01}$ can be 
obtained by the F-K analysis but not in general.   

The dependence of the velocities on $\alpha$  for a fixed $\beta$ is
shown in Fig.~{\ref{fig:2}}a.   
We see that three fronts move with different velocities for
$\alpha<\alpha_c{\left (\beta\right )}$ with 
$v_{01}>v_{-11}>v_{-10}$. 
All these velocities are same at $\alpha=\alpha_c{\left (\beta\right )}$.
For $\alpha>\alpha_c{\left (\beta\right )}$, the BZ front splits into two
fronts and the denatured state grows with time as
$\left (v_{-10}-v_{01}\right )t$.  It is straightforward to see that no 
stable front between $\pm 1$ can exist if $v_{-10}<v_{01}$.
Also the $v_{-11}$ curve ends at $\alpha_c{\left (\beta\right )}$ and has 
no continuation for $\alpha>\alpha_c{\left (\beta\right )}$.  This indicates 
that $\alpha_c{\left (\beta\right )}$ is a singular point. 
The numerically determined $\alpha_c{\left (\beta\right )}$ {\it vs}
$\beta$ line is shown in Fig.~{\ref{fig:3}}.  This is the {\it phase 
diagram for dynamics} with the phase boundary as the limit of stability 
of the BZ front(from below).

\phasedia 

The phase diagram can be confirmed by considering a few special cases.
For $\alpha=\beta$, the free energies of B and Z are same and the BZ front
should have zero velocity.  The point $\alpha=\beta=\frac{1}{\sqrt 3}$
corresponds to  the equilibrium situation, for which all the three fronts 
are static, and therefore the condition to be on the phase boundary is 
trivially satisfied.  This point is denoted by q in Fig.~{\ref{fig:3}}.   
Along the $\alpha=\beta$ line  for $\alpha<\alpha_c(\beta)$, 
$v_{01}, v_{-10}\neq0$ with state $+1$ or $-1$ 
invading $0$.  In contrast in  region 2, along the same 
$\alpha=\beta$ line, ``0'' is the stable state and it invades both 
$\pm 1$ states.  
In region 2 above the dotted line, obtained by equating 
$F(1)=F(0)$ (Eq.~(\ref{Eq:1})), the ``0'' state grows with the two 
fronts moving away from each other, but below that dotted line 
in region 1 the Z state grows though the fronts move in the 
same direction(towards right).  
The  Z $\leftrightarrow$ B symmetry in our choice of the free energy
mandates  a symmetric phase diagram across the $\alpha=\beta$ line
with the fronts moving towards left in region 3.   

For $\alpha, \beta$ close to $\alpha=\beta=\frac{1}{\sqrt 3}$, a 
perturbative analysis\cite{recG} can be done to determine the velocity, 
which is now a small parameter.  By writing, to first order in $v$,
\begin{equation}
  \label{eq:1}
U\left (z-vt\right )\approx U_0\left (z\right )+v\;t\; U_0^{\prime}\left 
(z\right ) 
\end{equation}
where prime denotes a derivative, $v$ can be determined to first order
in free energy difference if $U_0$ is known.  In the equilibrium 
situation, there is a Goldstone like zero-energy mode, because, the 
interface can be placed anywhere or shifted along $z$ without any 
cost of energy.  We therefore take $U_0(z)$ as centered around an 
arbitrarily chosen origin.
The static solution satisfies,
\begin{equation}
\frac{1}{2}\left (U_0^{\prime}(z)\right )^2=F\left (U_0\right
)=U_0^2\left (U_0^2-1\right )^2. 
\end{equation}
With a first order correction, the velocities are 
\begin{equation}\label{Eq:6}
v_{ij}=\frac{\epsilon_{ij}}{{\int_{-\infty}^\infty}[U_0^{\prime}(z)]^2dz},
\end{equation}
where $i,j=0,\pm 1,$ and the free energy differences
  $\epsilon_{ij}$ are
\begin{eqnarray}\label{Eq:7}
\epsilon_{01}&=&-\frac{1}{12}-2\frac{(\alpha-\beta)}{15}+\frac{\alpha\beta}{4
},\\
\epsilon_{-10}&=&\frac{1}{12}-2\frac{(\alpha-\beta)}{15}-\frac{\alpha\beta}{4
},\\
\epsilon_{-11}&=&-4\frac{(\alpha-\beta)}{15}.
\end{eqnarray}
At this perturbative regime, by equating the velocities, we find that 
around $\alpha=\beta=\frac{1}{\sqrt 3}$, the slope of the critical line 
is $-1$, which is consistent with the numerically determined boundary 
shown in Fig.~{\ref{fig:3}}.  Moreover we also find the phase boundary 
to deviate very slightly from a straight line over the range shown there.  
There is a deviation from linearity beyond that but the numerical error 
becomes larger. 
\width
We next study the behavior of the width of the interface and of the 
appropriate timescale for the dynamics.  For the special case of 
$\beta=0.5$ as $\alpha\rightarrow\alpha_c\left (\beta\right )$ the 
divergence of the width has been noted in Ref.~\cite{jhl}.  At 
$\alpha=\beta=\frac{1}{\sqrt 3}$, any length of ``$0$'' domain can be 
inserted and therefore the width of the BZ interface at the limit of 
stability is infinity.  
On the split-front side (Fig.~\ref{fig:2}d), the width increases 
linearly with time as $W=(v_{01}-v_{-1,0})t$ (Fig.~\ref{fig:4}a for 
$\alpha=0.75205$).  While, on the other side of the phase boundary 
the single front(Fig.~\ref{fig:2}b) has a finite width(Fig.~\ref{fig:4}a 
for $\alpha=0.745$).  Close to the phase boundary though a deformation of 
the moving front is visible around $\phi=0$(Fig.~\ref{fig:2}c), 
but width saturates at large time (Fig.~\ref{fig:4}a for $\alpha=0.75175$) 
without any appearance of the denatured phase.  Hence scaling forms are 
expected as
$$W\sim \mid \alpha-\alpha_c\left ((\beta\right )\mid^{-\mu},\  \and \ \tau\sim W^{\sf z}.$$
Fig.~\ref{fig:4}a shows the time evolution of the width of an interface for
various $\alpha$ at a fixed $\beta$, where the instantaneous width $W$ of the
interface at time $t$ is obtained  as
\begin{eqnarray}
  \label{eq:2}
  W^2&=&<z^2>-<z>^2, \ {\rm where\ } \\
 <z^n>&=&\frac{\int z^n \left(\frac{d\phi(z,t)}{dz}\right)^2 dz}{\int \left(\frac{d\phi(z,t)}{dz}\right)^2 dz}.
\end{eqnarray}
Another way to characterize the width would be to look at the slope of the 
profile {\it i.e.} $\left .\frac{d\phi(z)}{dz}\right |_{\phi=0}$, which is 
related to the inverse of $W$ and also shows the scaling with 
characteristic dynamic exponent.
We started with an interface that has an insertion of the ``0'' state and 
the width monitors the decay or the growth of the ``0'' state.  The width 
saturates exponentially for $\alpha<\alpha_c\left (\beta\right )$ albeit  
slowly near $\alpha\to\alpha_c-$, while a linear growth is observed for 
$\alpha>\alpha_c\left (\beta\right )$.  
Time here refers to the discretized time in the Crank-Nicolson approach.
By fitting an exponential to the time evolution of $W$, the characteristic 
time scale was determined, for $\alpha<\alpha_c\left (\beta\right )$.   
The exponent $\mu$ is found to be rather small, not inconsistent with the 
logarithmic growth observed in Ref. \cite{jhl}.
Fig.~\ref{fig:4}b shows the log-log plot of $\tau$ {\it vs} $W$ indicating 
a value of ${\sf z}$ within $3.0$ to $4.0$.  However for better accuracy 
one requires a large system and long time observation as well.   
The divergences of $W$ and $\tau$ with scaling establish the critical 
nature of the $\alpha=\alpha_c\left (\beta\right )$ line.    

Despite the immense success in probing the various phases of 
DNA by single molecular manipulation techniques, interfaces have not been 
explored thoroughly.
We hope our results will motivate direct studies of  interfaces of DNA, 
especially their stability.  Even on the theoretical front, it remains to 
be seen if all atom molecular dynamics simulations that have been
successful\cite{leeJPC,ji,ab} in seeing various phases, can be used to monitor 
the dynamics of interfaces, B-Z in particular,  under given boundary
conditions. 
\section{CONCLUSION}
The conformational transition from B-DNA to Z-DNA has been studied via
wave-front propagation.  The dynamic phase diagram for the  
steady state is 
obtained in the $\alpha$-$\beta$ plane, where $\alpha, \beta$ characterize 
the relative stability of the phases, by the critical value $\alpha_c$ for 
different values of $\beta$.  The phase boundary in the $\alpha$-$\beta$ 
plane has been determined and corroborated by a perturbation analysis.  
The dynamic transition is associated with diverging length and time scales 
and has its own dynamic exponent.  On one side of the phase boundary the 
dynamics involves propagation of one B-Z interface with a uniform speed, 
while on the other phase such an interface is unstable leading to the 
formation of the thermodynamically forbidden denatured state.  This in turn, 
suggests that there is no unique mechanism for the B-Z dynamics and it 
is possible to switch from one type to other by tuning the parameters.  
A resolution of the controversy in experiments is that the two cases, namely 
nanotube and magnetic tweezers are on the two sides of the phase boundary.

\end{document}